\documentclass[runningheads]{llncs}
\usepackage[T1]{fontenc}
\usepackage{amsmath,amssymb,amsfonts}
\usepackage{cancel}
\usepackage{graphicx}
\graphicspath{{figures/}}

\def\cN{\mathcal{N}}
\def\cF{\mathcal{F}}

\def\cJ{\mathcal{J}}
\def\-{\text{-}}
\def\+{\text{+}}
\def\tr{\text{tr}}
\newcommand\given[1][]{\:#1\vert\:}
\newcommand\euler{\textit{e}}

\usepackage[]{todonotes}

\usepackage{hyperref}
\usepackage{color}

\begin{document}
\title{Planning to avoid ambiguous states through Gaussian approximations to non-linear sensors\\ in active inference agents}
%
\titlerunning{Planning to avoid ambiguous states through Gaussian approximations}
%
\author{Wouter M. Kouw\orcidID{0000-0002-0547-4817}}
\authorrunning{W.M. Kouw}
\institute{Bayesian Intelligent Autonomous Systems laboratory\\
TU Eindhoven, Eindhoven, Netherlands \\
\email{w.m.kouw@tue.nl}
}
\maketitle              
\begin{abstract}
In nature, active inference agents must learn how observations of the world represent the state of the agent. In engineering, the physics behind sensors is often known reasonably accurately and measurement functions can be incorporated into generative models. When a measurement function is non-linear, the transformed variable is typically approximated with a Gaussian distribution to ensure tractable inference. We show that Gaussian approximations that are sensitive to the curvature of the measurement function, such as a second-order Taylor approximation, produce a state-dependent ambiguity term. This induces a preference over states, based on how accurately the state can be inferred from the observation. We demonstrate this preference with a robot navigation experiment where agents plan trajectories. 
\end{abstract}
\section{Introduction}
In nature, intelligent agents build a model to infer the causes of their sensations \cite{conant1970every}. In engineering, we are able to utilize knowledge of the relevant physics to structure such a model. In particular, we often know how sensors measure states of the world. For example, we know how radar measures relative velocity and distance \cite{zamiri2022bayesian}. Measurement functions that are non-linear transformations of state variables pose challenges to state estimation, which are often dealt with using Gaussian approximations of the transformed variables \cite{gustafsson2011some,sarkka2013bayesian}. We show that for certain Gaussian approximations, an active inference agent will prefer to avoid states because it already knows that state estimation will be difficult.

Active inference agents are based on free energy functionals that rank policies on explorative and goal-directed behaviour \cite{friston2010free,friston2015active,friston2017active,schwartenbeck2019computational}. The expected free energy functional can be understood through its decomposition into a cross-entropy term between states and observations given action ("ambiguity"), and a Kullback-Leibler divergence between the posterior predictive and a goal prior distribution ("risk") \cite{friston2015active,tschantz2020learning,da2020active}. We show that Gaussian approximations of a non-linear observation function that are itself linear in the covariance matrix, e.g., first-order Taylor and the unscented transform \cite{julier2004unscented}, lead to ambiguity terms that are constant over states. This echoes an earlier finding that agents with a linear Gaussian state-space model exhibit a constant ambiguity term \cite{koudahl2021epistemics}. However, utilizing a second-order Taylor approximation induces a non-constant ambiguity term. Under this model, the agent will avoid states where the non-linear measurement function curves strongly.
Our contributions are:
\begin{itemize}
    \item Analysis of ambiguity in expected free energy functions under three different Gaussian approximations.
    \item An experiment where a robot must plan a trajectory and navigate to a goal prior distribution, testing the effect of the ambiguity term.
\end{itemize}



\section{Problem statement}
We want to plan a trajectory for a robot across a plane. The robot's state at time $k$ is its planar position and time derivatives, $x_k \in \mathbb{R}^{D_x}$. The robot does not sense position directly, but has to infer it from noisy measurements $y_k \in \mathbb{R}^{D_y}$, produced by a sensor through a non-linear mapping $g : \mathbb{R}^{D_x} \rightarrow \mathbb{R}^{D_y}$ and measurement noise $v_k \in \mathcal{R}^{D_y}$. It accepts control inputs $u_k \in \mathbb{R}^{D_u}$ and moves according to linear dynamics with a transition matrix $A \in \mathbb{R}^{D_x \times D_x}$, control matrix $B \in \mathbb{R}^{D_x \times D_u}$ and process noise $e_k \in \mathbb{R}^{D_x}$. Overall, we consider robot systems described with discrete-time state-space models of the form:
\begin{align}
    x_k &= A x_{k-1} + B u_k + e_k \, , & e_k \sim \mathcal{N}(0, Q) \, , \\
    y_k &= g(x_k) + v_k \, , & v_k \sim \mathcal{N}(0, R) \, ,
\end{align}
where $Q,R$ are noise covariance matrices.

The goal is to find a sequence of $T$ controls $\bar{u}_k = u_{k+1}, \dots u_{k+T}$ that produces future states close to a desired state $x_{*}$. Agents must plan every time-step.
The challenge is that errors in state estimation may cause drastic changes in the planned trajectory, which can lead an agent astray. 

\subsubsection{Example} \label{sec:example} 
\begin{figure}
    \centering
    \includegraphics[width=.32\textwidth]{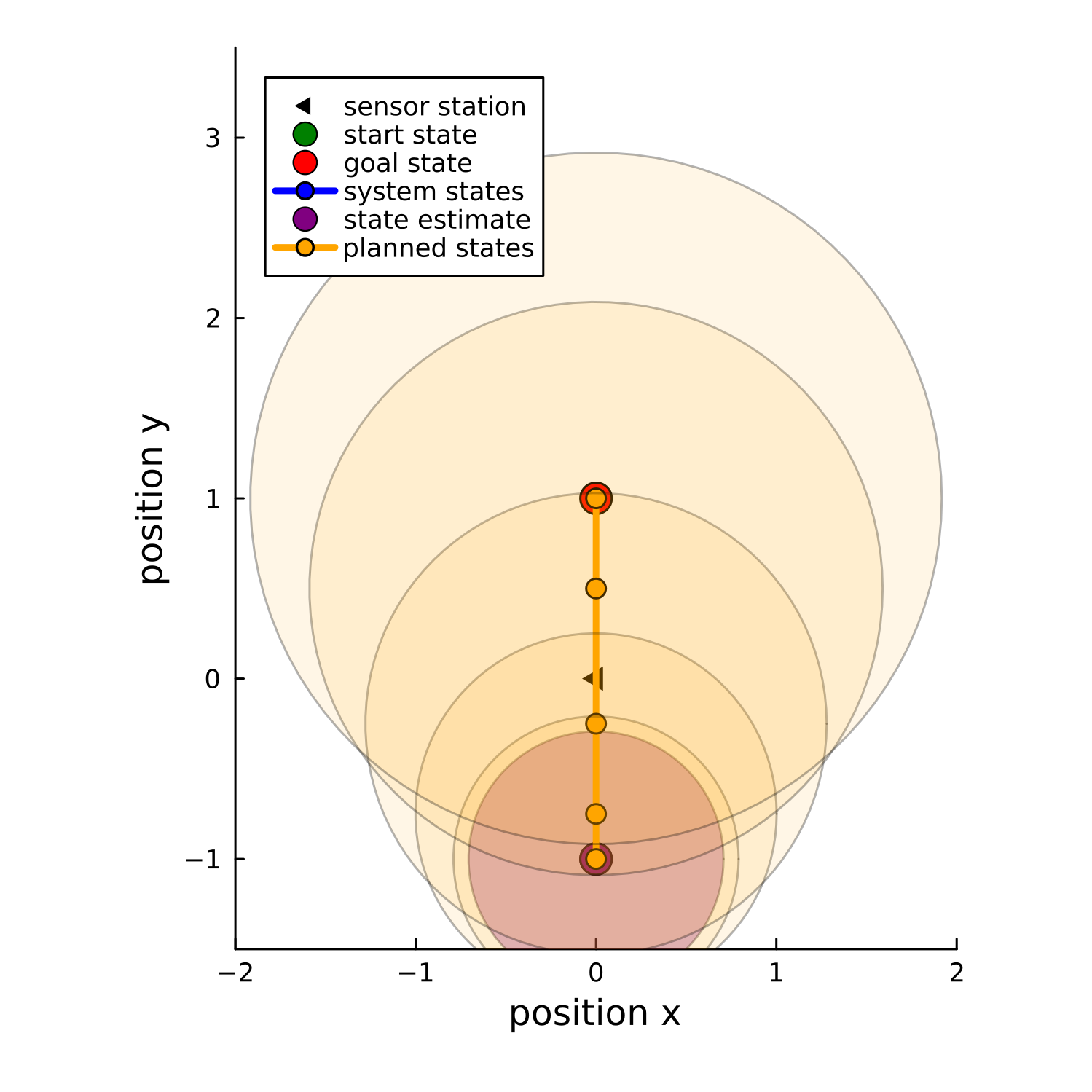}
    \includegraphics[width=.32\textwidth]{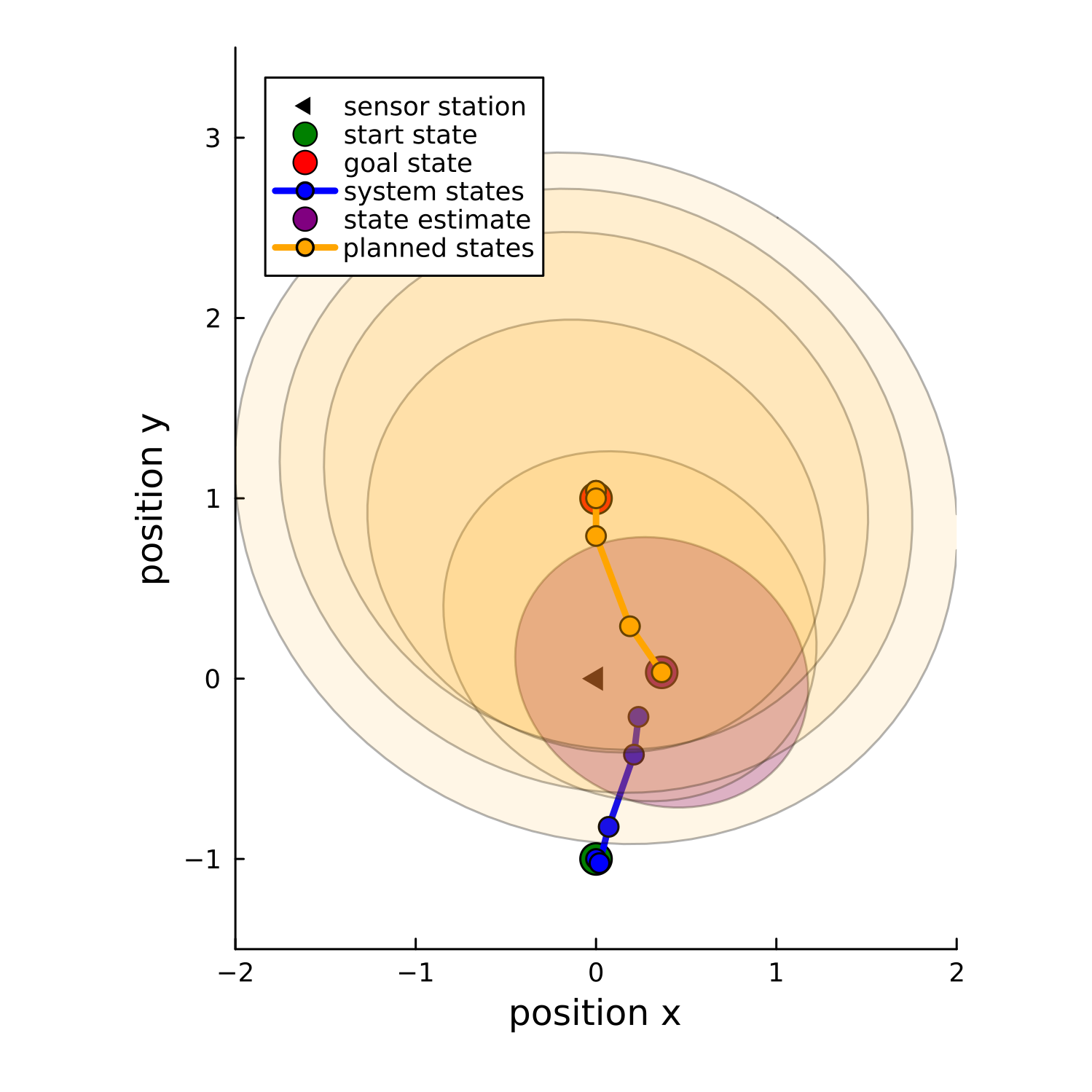}
    \includegraphics[width=.32\textwidth]{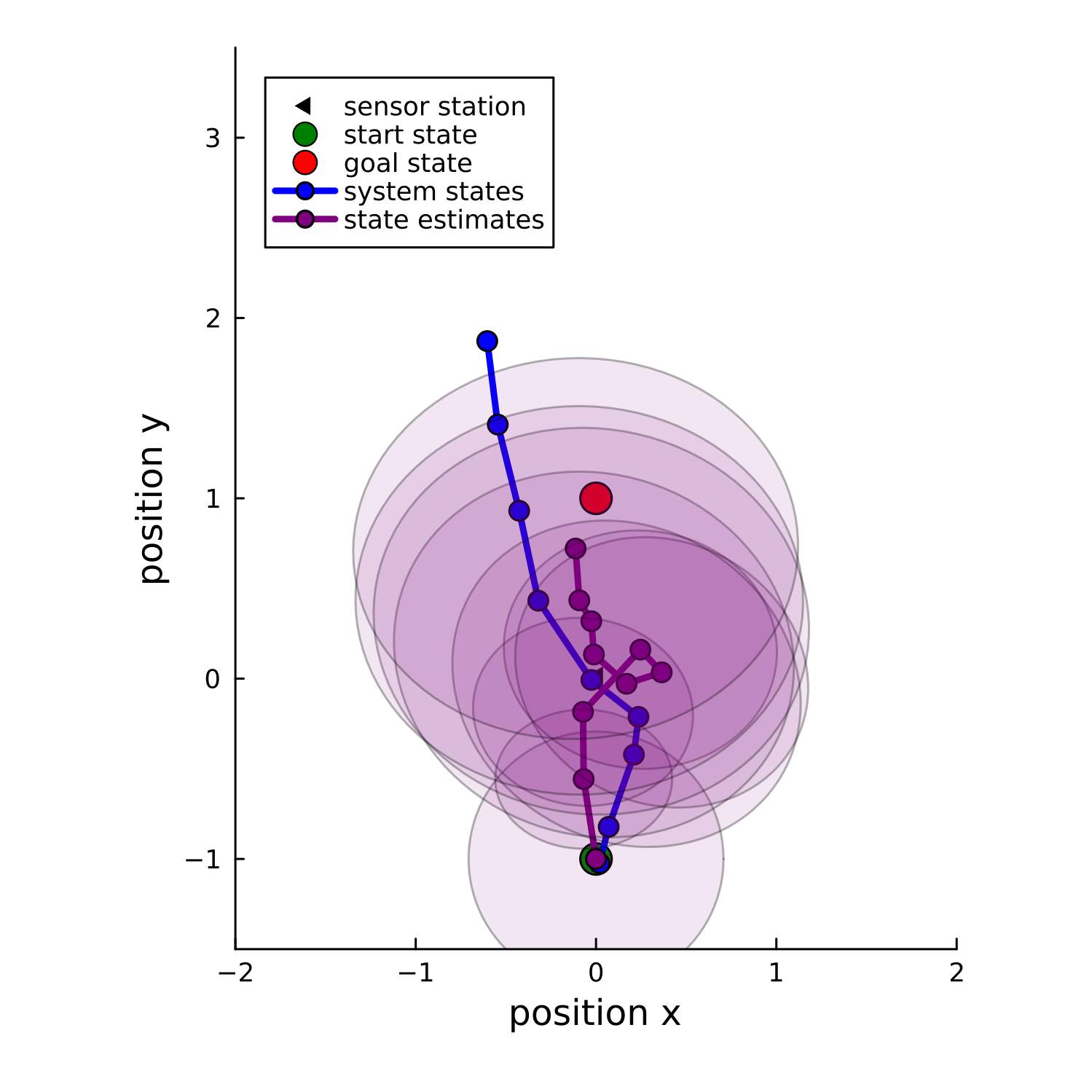}
    \caption{(Left) Planned trajectory at $k=1$, from start to goal directly over the sensor station. (Middle) Planned trajectory at $k=5$ showing a mismatch between true and estimated state resulting in a strong adjustment to the planned trajectory. (Right) Executed trajectory over a trial of $10$ steps demonstrates the agent losing track of the robot when it approaches the sensor station.}
    \label{fig:problem}
    \vspace{-20pt}
\end{figure}

Consider a robot with position and velocity states that must move from position $x_0 = (0, -1)$ to $x_{*} = (0, 1)$. Its state transition, control and process noise covariance matrices are given by:
\begin{align}
    A \! = \! \begin{bmatrix} 1 & 0 & \Delta t & 0 \\ 0 & 1 & 0 & \Delta t \\ 0 & 0 & 1 & 0 \\ 0 & 0 & 0 & 1 \end{bmatrix} , \, B \! = \! \begin{bmatrix} 0 & 0 \\ 0 & 0 \\ \Delta t & 0 \\ 0 & \Delta t \end{bmatrix} , \, Q \! = \! \begin{bmatrix} \sigma^2_1 \frac{\Delta t^3}{3} & 0 & \sigma^2_1 \frac{\Delta t^2}{2} & 0 \\ 0 & \sigma^2_2 \frac{\Delta t^3}{3} & 0 & \sigma^2_2 \frac{\Delta t^2}{2} \\ \sigma^2_1 \frac{\Delta t^2}{2} & 0 & \sigma^2_1 \Delta t & 0 \\ 0 & \sigma^2_2 \frac{\Delta t^2}{2} & 0 & \sigma^2_2 \Delta t   \end{bmatrix} ,
\end{align}
for $\Delta t = 0.5$, $\sigma_1 = \sigma_2 = 0.1$. Measurements are produced by a sensor station at $(0,0)$ that reports relative angle $\phi_k \in [-\pi \ \pi]$ and relative distance $d_k \in [0, \infty)$. The mapping and measurement noise covariance matrix are:
\begin{align}
    g(x_k) = \begin{bmatrix} \phi_k \\ d_k \end{bmatrix} = \begin{bmatrix} \sqrt{x_{1k}^2 + x_{2k}^2} \\
     \arctan(x_{1k}, x_{2k}) \end{bmatrix} \, , \ R =  \begin{bmatrix} \rho^2_1 & 0 \\ 0 & \rho^2_2 \end{bmatrix} \, ,
\end{align}
where $\rho_1 = \rho_2 = 0.001$.
Suppose it uses an extended Kalman filter (first-order Taylor approximation) for state estimation and a finite-horizon model-predictive control objective of the form:
\begin{align}
    J_k(\bar{u}_k) = \sum_{t=k+1}^{k+T} \big((A \hat{x}_{t-1} \! + \! Bu_t) - x_{*} \big)^{\intercal} C \big((A \hat{x}_{t-1} \! + \! Bu_t) - x_{*}\big) + \eta u_t^2 \, ,
\end{align}
where $\hat{x}$ is the mean state, $\hat{x}_t = A \hat{x}_{t-1} \! + \! Bu_t$, $C$ is a cost matrix (ones for position, zeros for velocity) and $\eta$ a regularization parameter. Minimizing this objective every time-step produces the control sequence $\bar{u}_k^{\text{MPC}} = \arg \min J_k(\bar{u}_k)$. Such an agent will first plan a trajectory moving directly forward, as described in Figure \ref{fig:problem} (left). However, as it approaches the sensor station, its state estimate become progressively more inaccurate and it makes increasingly more drastic adjustments to the control plan (see $k=5$ in Figure \ref{fig:problem} middle). Figure \ref{fig:problem} (right) shows the executed trajectory over a trial of 10 steps, demonstrating that the agent lost track of the robot's state and did not successfully reach the target.

\section{Agent specification}


\subsection{Probabilistic Model}
The agent's model will have Gaussian prior distributions over states and controls,
\begin{align}
    p(x_0) = \cN(x_0 \given m_0, S_0) \, , \qquad \cN(u_k \given 0, \eta^{-1} I) \label{eq:static-prior} \, ,
\end{align}
with mean $m_0$, covariance matrix $S_0$, precision $\eta$ and identity matrix $I$.
The agent's state transition will also be expressed as a Gaussian distribution:
\begin{equation}
    p(x_{k} \given x_{k-1}, u_{k}) = \mathcal{N}(x_{k} \given Ax_{k-1} + B u_{k}, Q) \, .
\end{equation}
%
Let the marginal state $x_k$ be Gaussian distributed, i.e., $p(x_k) = \cN(x_k \given m_k, S_k)$. We restrict our attention to approximations of the nonlinear sensor $g(x_k)$ that produce Gaussian joint distributions over states and observations \cite{sarkka2013bayesian}, i.e.,
\begin{align} \label{eq:gauss-approx-joint}
    p(y_k, x_k) \approx \cN \Big( \begin{bmatrix} x_k \\ y_k \end{bmatrix} \given \begin{bmatrix} m_k \\ \mu_k \end{bmatrix}, \begin{bmatrix} S_k & \Gamma_k \\ \Gamma_k^{\intercal} & \Sigma_k \end{bmatrix} \Big) \, .
\end{align}
From the joint, we obtain a conditional distribution of observations given states:
\begin{align} \label{eq:likelihood}
    p(y_k \given x_k) \approx \cN \big(y_k \given \mu_k + \Gamma_k^{\intercal} S_k^{-1}(x_k - m_k), \Sigma_k - \Gamma_k S_k^{-1} \Gamma_k^{\intercal} \big) \, .
\end{align}
This distribution is linear in $x_k$, and will allow for exact Bayesian filtering. But note that the parameters $\mu_k$, $\Gamma_k$ and $\Sigma_k$ may be nonlinear functions of $x_k$, depending on the type of Gaussian approximation (specifics treated in Section~\ref{sec:nln-obs}), and may thus capture more of the effect of $g(x_k)$. 
%
%

\subsection{Inferring states} \label{sec:fe-states}
We assume that, when inferring states, the agent has observed the system output $y_k = \hat{y}_k$ and input $u_{k}=\hat{u}_k$. 
Let $\mathcal{D}_k \triangleq \{\hat{y}_i, \hat{u}_i\}_{i=1}^{k}$ refer to data observed thus far.
Given the known executed control, state estimation follows the general Bayesian filtering equations \cite{sarkka2013bayesian}. Firstly, the prior predictive distribution is given by:
\begin{align}
    p(x_k | \, \hat{u}_k, \mathcal{D}_{k\-1}) = \int p(x_k | x_{k\-1}, \hat{u}_k) \, p(x_{k\-1} | \mathcal{D}_{k\-1}) \, \mathrm{d}x_{k\-1} = \mathcal{N}(x_k | \, \bar{m}_k, \bar{S}_k) \, . \label{eq:state-params}
\end{align}
with $\bar{m}_k \triangleq Am_{k-1} + B \hat{u}_{k}$ and $\bar{S}_k \triangleq A S_{k-1} A^{\intercal} + Q$. This prediction is corrected by the observation through Bayes' rule \cite{sarkka2013bayesian},
\begin{align}
    p(x_k \given \mathcal{D}_k) = \frac{p(\hat{y}_k \given x_k)}{p(\hat{y}_k \given \mathcal{D}_{k\-1})} p(x_k \given \hat{u}_k, \mathcal{D}_{k\-1}) = \mathcal{N}(x_k \given m_k, S_k) \, ,
\end{align}
with $m_k = \bar{m}_k + \Gamma_k \Sigma_k^{-1}(\hat{y}_k - \mu_k)$ and $S_k = \bar{S}_k - \Gamma_k \Sigma_k^{-1} \Gamma_k$. 

\subsection{Inferring controls} \label{sec:fe-controls}
We will discuss the inference procedure first for a single step into the future, and then generalize to a finite horizon of length $T$.
Predictions for the future state and observation are made by unrolling the generative model to $t=k+1$:
\begin{align} \label{eq:pmodel}
p(y_t,x_{t}, u_t \given \mathcal{D}_k) =  p(y_t \given x_t) p(x_t \given u_{t}; \mathcal{D}_k) p(u_{t}) \, .
\end{align}

We will use an expected free energy functional to infer a posterior distribution over the control $u_t$ \cite{millidge2021whence}:
\begin{align} \label{eq:efe-base}
\mathcal{F}_k[q] = \int q(y_t \given x_t) \int q(x_t, u_t) \ln \frac{q(x_t,u_t)}{p(y_t,x_t,u_t \given \mathcal{D}_k)} \mathrm{d}(u_t,x_t) \mathrm{d} y_t \, .
\end{align}
The variational model is specified to be:
\begin{align} \label{eq:qmodel}
    q(y_t \given x_t) \triangleq p(y_t \given x_t) \, , \quad  q(x_t, u_t) \triangleq p(x_t \given u_{t} ; \mathcal{D}_k) q(u_{t})  \, .
\end{align}
Constraining $q(y_t \given x_t)$ to the Gaussian approximation defined in Eq.~\ref{eq:likelihood} allows us to study deterministic approximations in an expected free energy minimization context.
%
Given this variational model, Eq.~\ref{eq:efe-base} may be re-arranged to: 
\begin{align}
\cF_k[q] 
&= \int p(y_t \given x_t) \int p(x_t \given u_t; \mathcal{D}_{k}) q(u_t) \ln \frac{p(x_t \given u_t; \mathcal{D}_{k}) q(u_t)}{p(y_t,x_t,u_t; \mathcal{D}_k)} \mathrm{d}(x_t, u_t) \mathrm{d}y_t \\
&= \! \! \int \! q(u_t) \big(\int p(y_t,x_t \given u_t; \mathcal{D}_k) \ln \frac{p(x_t \given u_t; \mathcal{D}_k) q(u_t)}{p(y_t,x_t \given u_t; \mathcal{D}_k) p(u_t)} \mathrm{d}(y_t,x_t) \big) \mathrm{d}u_t  \\
&= \! \! \int \! q(u_t) \big(\! \ln \! \frac{q(u_t)}{p(u_t)} \! + \! \! \underbrace{ \int \! p(y_t,x_t | u_t; \mathcal{D}_k) \ln \! \frac{p(x_t \given u_t; \mathcal{D}_k)}{p(y_t,x_t | u_t; \mathcal{D}_k)} \mathrm{d}(y_t, x_t) }_{\cJ_k(u_t)} \! \big) \mathrm{d}u_t . \label{eq:FE2}
\end{align}
We refer to $\cJ_k(u_t)$ as the expected free energy \emph{function} as it depends on the value of $u_t$ not on its distribution.
Under $\cJ_k(u_t) = \ln (1/ \exp(-\cJ_k(u_t)))$, the expected free energy functional can be concisely expressed as:
\begin{align}
\cF_k[q] &= \int q(u_t) \ln \frac{q(u_t)}{p(u_t)\exp\left(- \cJ_k(u_t)\right) } \mathrm{d}u_t 
\, .
\end{align}
The above is a Kullback-Leibler divergence, which is minimal when 
 \begin{align}
     q^{*}(u_t) \propto p(u_t)\exp\left(- \cJ_k(u_t)\right) \, .
 \end{align}
The proportionality is due to the implicit constraint\footnote{A more rigorous treatment would define a Lagrangian with normalization and marginalization constraints \cite{senoz2021variational}. However, such a treatment is inconsequential when resorting to MAP estimation, as will be pursued later in the paper.} that $q^{*}(u_t)$ should integrate to $1$.
%
%
To work out the expectation in Eq.~\ref{eq:FE2}, we first decompose the joint over states and observations into
\begin{align} \label{eq:efe-steps02}
    p(y_t,x_t \given u_t; \mathcal{D}_k) = p(x_t \given y_t, u_t; \mathcal{D}_k) p(y_t) \, ,
\end{align}
and then intervene on the marginal distribution over $y_t$ with a  distribution reflecting desired future observations (a.k.a. goal prior) \cite{van2022active}: 
\begin{align} \label{eq:goalprior}
    p(y_t) \rightarrow p(y_t \given y_{*}) = \cN(y_t \given \mu_{*}, \Sigma_{*}) \, .
\end{align}
%
The next step involves applying Bayes' rule in the inverse direction:
\begin{align} \label{eq:efe-bayesrule}
    \frac{1}{p(x_t \given y_t, u_t; \mathcal{D}_k)} = \frac{p(y_t \given u_t; \mathcal{D}_k)}{p(y_t \given x_t)p(x_t \given u_t; \mathcal{D}_k)} \, ,
\end{align}
where the marginal prediction for the future observation is:
\begin{align} 
    p(y_t \given u_t; \mathcal{D}_k) &= \int p(y_t \given x_t) \, p(x_t \given u_t; \mathcal{D}_k) \mathrm{d}x_t \\
    &= \int \mathcal{N}(\begin{bmatrix} x_t \\ y_t \end{bmatrix} \given \begin{bmatrix}  \bar{m}_{t}  \\ \mu_t \end{bmatrix}, \begin{bmatrix} \bar{S_t} &  \Gamma_t \\ \Gamma_t^{\intercal} & \Sigma_t \end{bmatrix}) \mathrm{d}x_t = \mathcal{N}(y_t \given \mu_t, \Sigma_t) \, . \label{eq:postpred}
\end{align}
Note that $\mu_t$ and $\Sigma_t$ depend on $u_t$ through $\bar{m}_t$.
Plugging Eqs.~\ref{eq:goalprior} and \ref{eq:efe-bayesrule} into Eq.~\ref{eq:efe-steps02} yields:
\begin{align}
 \cJ_k(u_t) 
 \! &= \! \int \! p(y_t, x_t | u_t; \mathcal{D}_k) \ln \frac{p(x_t | u_t; \mathcal{D}_k)}{p(y_t \given y_{*})} \frac{p(y_t \given u_t; \mathcal{D}_k)}{p(y_t \given x_t)p(x_t \given u_t; \mathcal{D}_k)} \, \mathrm{d}(y_t, x_t) \\
&= \! \underbrace{\int \! p(y_t, x_t \given u_t; \mathcal{D}_k) \Big[ -\ln \frac{p(y_t, x_t \given u_t; \mathcal{D}_k)}{p(x_t \given u_t; \mathcal{D}_k)} \Big] \mathrm{d}(y_t,x_t)}_{\text{ambiguity}} \nonumber \\
&\qquad \qquad + \underbrace{\int \Big[ \int p(y_t, x_t \given u_t; \mathcal{D}_k) \mathrm{d}x_t \Big] \ln \frac{p(y_t \given u_t; \mathcal{D}_k)}{p(y_t \given y_{*})} \, \mathrm{d}y_t}_{\text{risk}} \label{eq:EFE-ambrisk}.
\end{align}
"Risk" refers to the Kullback-Leibler (KL) divergence between predicted and desired future observations. The inner integral in the risk term leads to a Gaussian distribution (Eq.~\ref{eq:postpred}) and  
the KL divergence between Gaussians is \cite{cover1999elements}:
\begin{align} \label{eq:risk}
\mathbb{E}_{p(y_t \given u_t; \mathcal{D}_k)} \Big[ \ln \frac{p(y_t \given u_t; \mathcal{D}_k)}{p(y_t \given y_{*})} \Big]  =  \frac{1}{2} \Big( \ln\frac{|\Sigma_{*}|}{|\Sigma_t |}  - D_y  + \text{tr}\big(\Sigma_{*}^{-1} (\Sigma_t + \Psi_{*}) \big)  \Big) \, .
\end{align}
where $\Psi_{*} \triangleq (\mu_{*}  -  \mu_t \big)\big(\mu_{*}  -  \mu_t)^{\intercal}$.
"Ambiguity" refers to the conditional entropy of the future observations given the future states.
\begin{lemma} \label{lemma1}
Ambiguity, as defined in Eq.~\ref{eq:EFE-ambrisk}, for a generative model described in Eq.~\ref{eq:pmodel} and a variational distribution described in Eq.~\ref{eq:qmodel}, is:
\begin{align} \label{eq:ambiguity}
    \mathbb{E}_{p(y_t, x_t | u_t; \mathcal{D}_k)} \big[ \! - \! \ln \frac{p(y_t, x_t | u_t; \mathcal{D}_k)}{p(x_t | u_t; \mathcal{D}_k)} \big] = \frac{D_y}{2} \ln (2\pi\euler) + \frac{1}{2} \ln|\Sigma_t - \Gamma^{\intercal}_t \bar{S}_t^{\- 1} \Gamma_t |  .
\end{align}
\end{lemma}
The proof is in Appendix \ref{app:A}. Note that the first term does not depend on the state $x_t$.
Plugging Eqs.~\ref{eq:risk} and \ref{eq:ambiguity} into the expected free energy function (Eq.~\ref{eq:EFE-ambrisk}) produces:
\begin{align} \label{eq:Ju}
    \cJ_k(u_t) \! = \! \frac{1}{2} \Big(\ln\frac{|\Sigma_{*}|}{|\Sigma_t |} \! + \! D_y \ln (2\pi) \! + \! \text{tr}\big(\Sigma_{*}^{-1} ( \Sigma_t \! + \! \Psi_{*} ) \big) \! + \! \ln \big|\Sigma_t \! - \! \Gamma^{\intercal}_t \bar{S}_t^{\- 1} \Gamma_t \big|  \Big) \, .
\end{align}
Note that $\Gamma_t$, $\Sigma_t$ and $\Psi_{*}$ depend on $u_t$.
The above steps can be generalized to a longer time horizon $t = k+1,  \dots k+T$.
Because the prior is independent over time, $p(\bar{u}) = \prod_{t=1}^{T} p(u_t)$ (see Eq.~\ref{eq:static-prior}), the function $\cJ_k(\bar{u})$ factorizes to a sum of recursive expected free energy functions $\sum_{t=1}^{T} \cJ_k(u_t)$.

We are interested in the most probable value under the approximate control posterior, i.e., the MAP estimate:
\begin{align}
    \hat{u} &= \underset{\bar{u} \in \mathcal{U}}{\arg \max} \ q^{*}(\bar{u}) = \underset{\bar{u} \in \mathcal{U}}{\arg \min} \sum_{t=k+1}^{k+T} \cJ_t(u_t) - \ln p(u_t)  \, ,
\end{align}
where $\mathcal{U} \subset \mathbb{R}^{T}$ is the space of affordable controls over $T$ steps. Constraints such as motor force limits can be imposed during optimization.

\section{Gaussian approximations} \label{sec:nln-obs}
We discuss the three most popular Gaussian approximations to non-linear transformations of Gaussian random variables: the first and second-order Taylor series approximations (used in extended Kalman filters) and the unscented transform (used in the unscented Kalman filter) \cite{julier2004unscented,gustafsson2011some}\cite[Ch.~5]{sarkka2013bayesian}.

The first-order Taylor series approximation effectively linearizes the non-linear observation function $g(x_t)$. Since ambiguity is known to be constant over states under a linear observation function \cite{koudahl2021epistemics}, it is no surprise that the first-order Taylor also leads to an ambiguity term that is constant over states.
\begin{theorem} \label{th:01}
Let $G_x(\bar{m}_t)$ be the Jacobian of $g$ with respect to $x_t$, evaluated at $\bar{m}_t$. Under a \emph{first}-order Taylor approximation, the parameters $\Sigma_t, \Gamma_t$ are:
\begin{align} \label{eq:taylor1}
    \Sigma_t = G_x(\bar{m}_t) \bar{S}_t G_x(\bar{m}_t)^{\intercal} + R \, , \qquad 
    \Gamma_t = \bar{S}_t G_x(\bar{m}_t)^{\intercal} \, .
\end{align}
With these parameters, the ambiguity term does not depend on the state $x_t$:
\begin{align}
    \mathbb{E}_{p(y_t, x_t \given u_t; \mathcal{D}_k)} \big[ -\ln \frac{p(y_t, x_t \given u_t; \mathcal{D}_k)}{p(x_t \given u_t; \mathcal{D}_k)} \big] = \frac{D_y}{2} \ln (2\pi\euler) + \frac{1}{2}\ln | R | \, .
\end{align}
\end{theorem}
The proof is in Appendix \ref{app:B}. 
Perhaps surprisingly, under the second-order Taylor approximation, the ambiguity term varies as a function of the state $x_t$.
\begin{theorem} \label{th:02}
Let $G_{xx}^{(i)}(\bar{m}_t)$ be the Hessian of the $i$-th element of the non-linear observation function evaluated at $\bar{m}_t$ and $e_i$ be a canonical basis vector. The parameters $\Sigma_t, \Gamma_t$ computed through a \emph{second}-order Taylor approximation are:
\begin{align}
    \Sigma_t &= G_x(\bar{m}_t) \bar{S}_t G_x(\bar{m}_t)^{\intercal} +  \frac{1}{2} \sum_{i=1}^{D_y} \sum_{j=1}^{D_y} e_i e_j^{\intercal} \tr \big(G_{xx}^{(i)}(\bar{m}_t) \bar{S}_t G_{xx}^{(j)}(\bar{m}_t) \bar{S}_t \big) + R \nonumber \\ 
    \Gamma_t &= \bar{S}_t G_x(\bar{m}_t)^{\intercal} \, . \label{eq:taylor2}
\end{align}
With these parameters, the ambiguity term depends on $x_t$ through:
\begin{align}
    &\mathbb{E}_{p(y_t, x_t \given u_t;  \mathcal{D}_k)}  \big[ -\ln   \frac{p(y_t, x_t \given u_t; \mathcal{D}_k)}{p(x_t \given u_t;  \mathcal{D}_k)} \big] = \\
    &\qquad \frac{D_y}{2} \ln (2\pi\euler) +  \frac{1}{2}  \ln \Big| \frac{1}{2} \sum_{i=1}^{D_y} \sum_{j=1}^{D_y} e_i e_j^{\intercal} \text{tr} \big( G_{xx}^{(i)}(\bar{m}_t) \bar{S}_t G_{xx}^{(j)}(\bar{m}_t) \bar{S}_t \big)  +  R \, \Big|  \, . \nonumber 
\end{align}
\end{theorem}
The proof is in Appendix \ref{app:C}.

\paragraph{} Interestingly, the ambiguity is also constant for the unscented transform.
%
\begin{theorem} \label{th:03}
Define $2D_x + 1$ sigma points as:
\begin{align} \label{eq:ut_spoints}
    \chi_0 \triangleq \bar{m}_t , \ \ \chi_i \triangleq \bar{m}_t \! + \! \sqrt{D_x + \lambda} \big[ \sqrt{\bar{S}_t} \big]_i , \ \ \chi_{D_x+i} \triangleq \bar{m}_t \! - \! \sqrt{D_x + \lambda} \big[ \sqrt{\bar{S}_t} \big]_i ,
\end{align}
where $i = 1, \dots D_x$, $[ \cdot ]_i$ denotes the $i$-th column of a matrix, and $\sqrt{S}$ denotes the matrix square root such that $\sqrt{S} \sqrt{S} = S$. The parameter $\lambda \triangleq \alpha^2 (D_x + \kappa) - D_x$ depends on free parameters $\alpha$ and $\kappa$. Define $2D_x + 1$ weights as:
\begin{align} \label{eq:ut_weights}
    w_0 \triangleq \frac{\lambda}{D_x + \lambda} + (1-\alpha^2+\beta) \ , \qquad w_i \triangleq \frac{1}{D_x + \lambda} \, ,
\end{align}
for $i = 1, \dots 2D_x$ and $\beta$ as an additional free parameter.
Under these sigma points and weights, the parameters $\mu_t$, $\Sigma_t$, $\Gamma_t$ are \cite[Eq.~5.89]{sarkka2013bayesian}:
\begin{align} \label{eq:ut-sigma-gamma}
    \mu_t &= \frac{\lambda}{D_x \! + \! \lambda} g(\chi_0) + \sum_{i=1}^{2D_x} \frac{1}{2(D_x \! + \! \lambda)} g(\chi_i) \, , \quad
    \Gamma_t  =  \sum_{i=0}^{2D_x} w_i (\chi_i \! - \! \bar{m}_t)(g(\chi_i) \! - \! \mu_t)^{\intercal} \, ,\nonumber \\
    &\qquad \Sigma_t =  \sum_{i=0}^{2D_x} w_i (g(\chi_i)  -  \mu_t)(g(\chi_i)  -  \mu_t)^{\intercal}  +  R \, .
\end{align}
Then, the ambiguity is independent of the state:
\begin{align}
    \mathbb{E}_{p(y_t, x_t \given u_t;  \mathcal{D}_k)} & \big[ -\ln \frac{p(y_t, x_t \given u_t; \mathcal{D}_k)}{p(x_t \given u_t;  \mathcal{D}_k)} \big] = \frac{D_y}{2} \ln (2\pi\euler) + \frac{1}{2} \ln |R|  \, .
\end{align}
\end{theorem}
The proof can be found in the Appendix \ref{app:D}. This result is conjectured to hold for other Gaussian approximations that are linear in their estimate of the covariance matrix, for example the Gauss-Hermite approximation \cite[Ch.~6]{sarkka2013bayesian}.

\section{Experiments}

Our experiment is as described in Section \ref{sec:example}, with the nonlinear observation function $g(\cdot)$ measuring relative angle and distance to a base station. Examples of sensors include Hall effect and ultrasound sensors. The robot starts at $x_0 = [0 \ \text{-}1 \ 0 \ 0]$ and must reach $x_{*} = [0 \ 1 \ 0 \ 0]$. The agent's state prior distribution's parameters were $m_0 = [0 \ \text{-}1 \ 0 \ 0]$ and $S_0 = 0.5 I$. Its control prior precision was set to a tiny value, $\eta = 1.0\cdot 10^{-8}$, so as to best study the effects of ambiguity and risk. It was given a goal prior of $m_{*} = g(x_{*})$ and $S_{*} = 0.5 I$.

\begin{figure}
    \includegraphics[width=\textwidth]{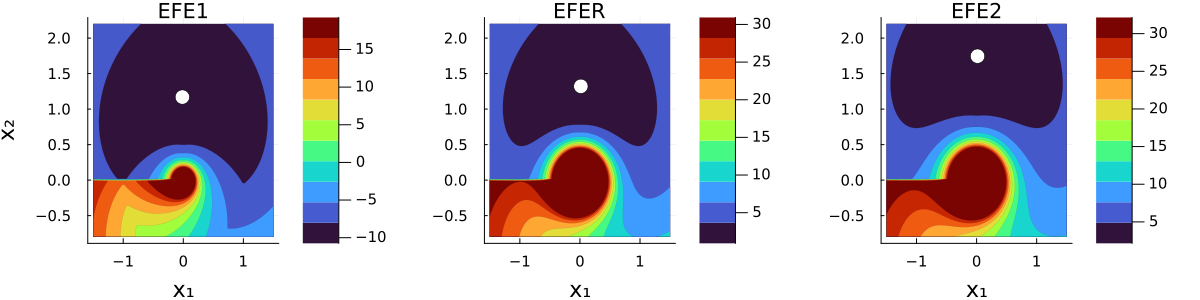}
    \caption{Value under three EFE functions over a plane: EFE1 is risk and ambiguity under a first-order Taylor approximation, EFER is risk only under a second-order Taylor approximation and EFE2 is both risk and ambiguity under a second-order Taylor approximation. White markers indicate minimizers. Note that each EFE function induces a different preference over states.}
    \label{fig:ambiguity}
    \vspace{-10pt}
\end{figure}

We will compare three agents\footnote{Details and code at: \href{https://github.com/biaslab/IWAI2024-ambiguity}{https://github.com/biaslab/IWAI2024-ambiguity}}: firstly, an agent that uses the first-order Taylor approximation, referred to as EFE1. Secondly, an agent with a second-order Taylor approximation, referred to as EFE2. Thirdly, an agent with a second-order Taylor approximation but with only the risk term included, referred to as EFER. The difference between EFER and EFE2 reflects the effect of the ambiguity term, while the difference between EFE1 and EFE2 reflects the effect of the second-order Gaussian approximation. Figure \ref{fig:ambiguity} plots the value of the control objective function at every position in state-space, under a state covariance matrix of $S_t=I$. States close to the sensor station are red and will lead to high values under the control objective. Note that the area around the sensor station increases from EFE1 to EFER due to the curvature of the relative distance sensor. The white markers are the approximate minimizers for this choice of $S_t$ matrix. Comparing EFER and EFE2, we can see that ambiguity increases the cost of being close to the sensor station. 

\begin{figure}
    \includegraphics[width=.32\textwidth]{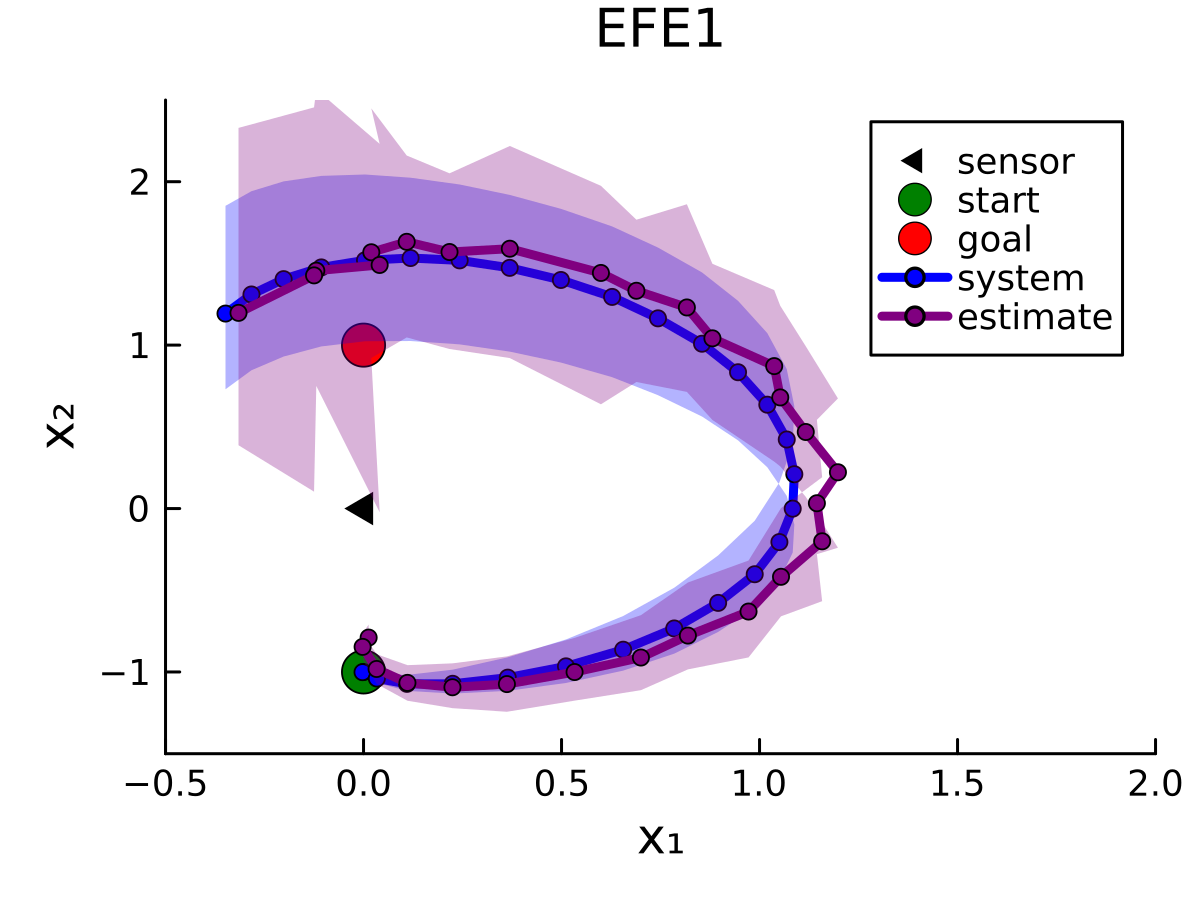}
    \includegraphics[width=.32\textwidth]{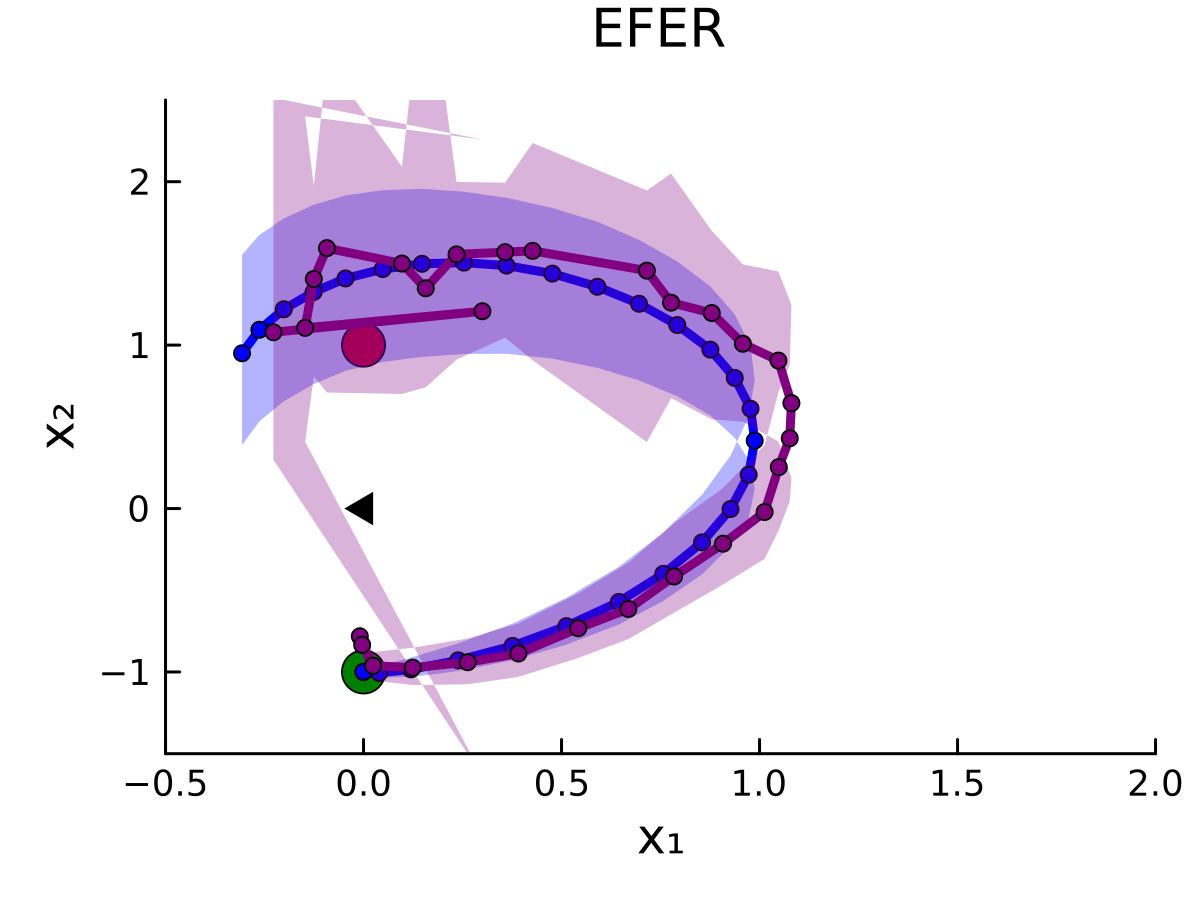}
    \includegraphics[width=.32\textwidth]{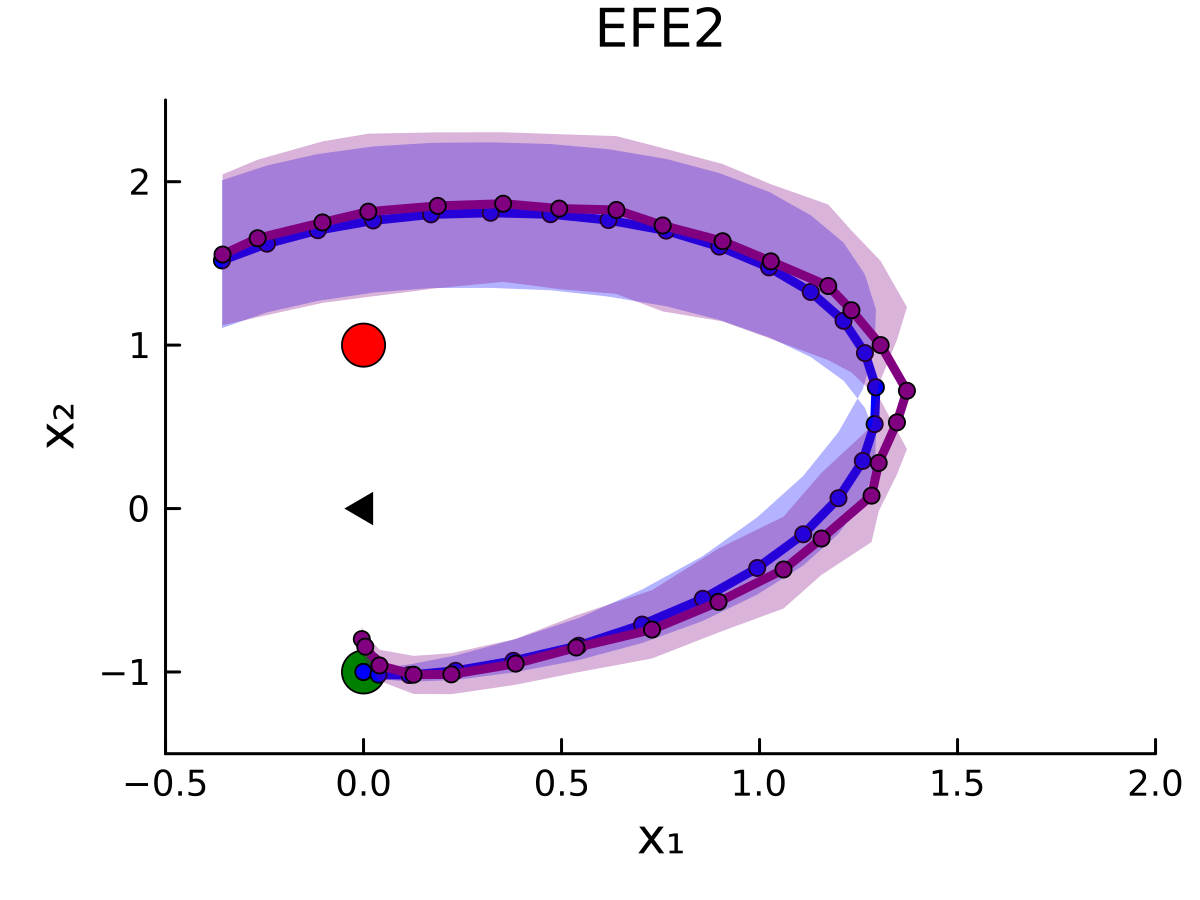}
    \caption{Trajectories of agents under three EFE functions, averaged over 100 Monte Carlo samples (ribbon is standard deviation of the mean). The robot starts at the green marker and must reach the red goal marker. All agents avoid the sensor station, with EFE2 taking the widest curve and having the smoothest average trajectory.}
    \label{fig:paths}
\end{figure}

We ran 100 Monte Carlo experiments. Figure \ref{fig:paths} plots the average trajectory of $T=30$ steps taken by the EFE1, EFER and EFE2 agents. Ribbons indicate the standard error of the mean at every time-point. Note that all agents avoid the sensor station, with EFE2 taking the widest curve (EFE1 and EFER turn at $x_1 = 1.0$ while EFE2 turns at $x_1 = 1.5$). EFE1 and EFER lose track of the robot in a number of experiments (like the model predictive controller in Sec.~\ref{sec:example}), leading to a more volatile average trajectory. EFE2 has the smoothest average trajectory, indicating that the ambiguity term helps planning.

\section{Discussion} \label{sec:discussion}

One could argue that our analysis is more about model selection than inference, as each Gaussian approximation essentially constitutes a different generative model. 
In that sense, the experiments only indicate that richer approximations of nonlinear functions lead to better performance, which is not surprising. 
However, the result is more subtle than that since the unscented transform is richer than the first-order Taylor (produces a more accurate mean estimate \cite{gustafsson2011some}) but apparently still leads to constant ambiguity. No, the approximation must be sensitive to how the covariance matrix of the joint distribution over states and observations changes as a function of $g$'s curvature. It would be interesting to extend this work with parameter estimation, such as inferring the process noise covariance matrix using a Wishart distribution \cite{sarkka2009recursive}, or the state transition matrix with a Matrix-Normal distribution \cite{barber2006unified,luttinen2013fast}. 




%




\section{Conclusion}
We examined active inference agents with linear Gaussian distributed dynamics and a non-linear measurement function. We found that the first-order Taylor series and unscented transform approximations to the non-linearly transformed states lead to expected free energy functions with ambiguity terms that are constant over states. A second-order Taylor approximation leads to a state-dependent ambiguity term, inducing a preference over states. 

\begin{credits}
\subsubsection{\ackname} The author gratefully acknowledges financial support from the Eindhoven Artificial Intelligence Systems Institute (EAISI) at TU Eindhoven.

\subsubsection{\discintname}
The authors have no competing interests to declare that are
relevant to the content of this article.
\end{credits}

\appendix

\section{Appendix: proof of Lemma 1} \label{app:A}
\begin{proof}
The cross-entropy is split into two entropies that simplify according to:
    \begin{align}
    &\mathbb{E}_{p(y_t, x_t \given u_t; \mathcal{D}_k)} \big[ -\ln \frac{p(y_t, x_t \given u_t; \mathcal{D}_k)}{p(x_t \given u_t; \mathcal{D}_k)} \big] \nonumber \\
    &= - \int  \mathcal{N}\big(\begin{bmatrix} x_t \\ y_t \end{bmatrix} \given \begin{bmatrix}  \bar{m}_{t}  \\ \mu_t \end{bmatrix}, \begin{bmatrix} \bar{S_t} &  \Gamma_t \\ \Gamma_t^{\intercal} & \Sigma_t \end{bmatrix}\big) \ln \mathcal{N}\big(\begin{bmatrix} x_t \\ y_t \end{bmatrix} \given \begin{bmatrix}  \bar{m}_{t}  \\ \mu_t \end{bmatrix}, \begin{bmatrix} \bar{S_t} &  \Gamma_t \\ \Gamma_t^{\intercal} & \Sigma_t \end{bmatrix}\big) \, \mathrm{d}(y_t,x_t)  \nonumber \\
    &\qquad + \int  \mathcal{N}(x_t \given \bar{m}_t, \bar{S}_t) \ln \mathcal{N}(x_t \given \bar{m}_t, \bar{S}_t) \mathrm{d}x_t  \\
    &= \frac{D_x+D_y}{2}\ln (2\pi \euler) + \frac{1}{2}\ln \big|\begin{bmatrix} \bar{S_t} &  \Gamma_t \\ \Gamma^{\intercal}_t & \Sigma_t \end{bmatrix} \big| - \Big(\frac{D_x}{2} \ln (2\pi \euler) + \frac{1}{2} \ln| \bar{S}_t| \Big)  \\
    &= \frac{D_y}{2} \ln(2\pi\euler) + \frac{1}{2}\ln\big( |\bar{S_t}| \cdot |\Sigma_t - \Gamma^{\intercal}_t \bar{S}^{-1}_t \Gamma_t | \big)  - \frac{1}{2} \ln| \bar{S}_t| \\
    &= \frac{D_y}{2} \ln (2\pi\euler) + \frac{1}{2} \ln|\Sigma_t - \Gamma^{\intercal}_t \bar{S}_t^{\- 1} \Gamma_t | \, .
\end{align}
\end{proof}

\section{Appendix: proof of Theorem 1} \label{app:B}
\begin{proof}
    Plugging $\Sigma_t$, $\Gamma_t$ from \eqref{eq:taylor1} into the result from Lemma \ref{lemma1}, yields:
    \begin{align}
        &\frac{D_y}{2} \ln (2\pi\euler) + \frac{1}{2}\ln | \Sigma_t - \Gamma_t^{\intercal} \bar{S}_t^{-1} \Gamma |  \\ 
        &\quad = \frac{D_y}{2} \ln (2\pi\euler) + \frac{1}{2}\ln | G_x(\bar{m}_t) \bar{S}_t G_x(\bar{m}_t)^{\intercal} + R - G_x(\bar{m}_t) \bar{S}_t^{\intercal} \bar{S}_t^{-1} \bar{S}_t G_x(\bar{m}_t)^{\intercal} |  \nonumber \\
        &\quad = \frac{D_y}{2} \ln (2\pi\euler) + \frac{1}{2} \ln |R| \, .
    \end{align}
    The cancellation is due to $\bar{S}_t$ being symmetric, i.e., $\bar{S}_t^{\intercal}\bar{S}_t^{-1} = \bar{S}_t \bar{S}_t^{-1} = I$.
\end{proof}

\section{Appendix: proof of Theorem 2} \label{app:C}
\begin{proof}
    Plugging $\Sigma_t$, $\Gamma_t$ from \eqref{eq:taylor2} into the result from Lemma \ref{lemma1}, yields:
    \begin{align}
        &\frac{D_y}{2} \ln (2\pi\euler) \! + \! \frac{1}{2}\ln \Big| \, \Sigma_t \! - \! \Gamma_t^{\intercal} \bar{S}_t^{-1} \Gamma \, \Big| = \! \frac{D_y}{2} \ln (2\pi\euler) \! + \! \frac{1}{2}\ln \Big| \, G_x(\bar{m}_t) \bar{S}_t G_x(\bar{m}_t)^{\intercal}  \\
        & \quad + \frac{1}{2} \sum_{i=1}^{D_y} \sum_{j=1}^{D_y} e_i e_j^{\intercal} \tr \big(G_{xx}^{(i)}(\bar{m}_t) \bar{S}_t G_{xx}^{(j)}(\bar{m}_t) \bar{S}_t \big) + R -  G_x(\bar{m}_t) \bar{S}_t^{\intercal} \bar{S}_t^{-1} \bar{S}_t G_x(\bar{m}_t)^{\intercal} \Big| \nonumber \\
        &\ = \frac{D_y}{2} \ln (2\pi\euler) + \frac{1}{2} \ln \Big| \, \frac{1}{2} \sum_{i=1}^{D_y} \sum_{j=1}^{D_y} e_i e_j^{\intercal} \tr \big(G_{xx}^{(i)}(\bar{m}_t) \bar{S}_t G_{xx}^{(j)}(\bar{m}_t) \bar{S}_t \big) + R \, \Big| \, .
    \end{align}
    The covariance matrix $\bar{S}_t$ is symmetric, i.e., $\bar{S}_t^{\intercal}\bar{S}_t^{-1} = \bar{S}_t\bar{S}_t^{-1} = I$. 
    Note that the Hessian $G_{xx}^{(i)}(\bar{m}_t)$ depends on the inferred mean of the predicted state $\bar{m}_t$, meaning that ambiguity is not constant over state-space.
\end{proof}

\section{Appendix: proof of Theorem 3} \label{app:D}
\begin{proof}
    Plugging $\mu_t$, $\Sigma_t$, $\Gamma_t$ from \eqref{eq:ut-sigma-gamma} into the log-determinant term from the result in Lemma \ref{lemma1}, gives:
    \begin{align} \label{eq:proof_ut01}
        & \frac{1}{2} \ln \big| \Sigma_t - \Gamma_t^{\intercal} \bar{S}_t^{-1} \Gamma \big| = \frac{1}{2} \ln \Big| \sum_{i'=0}^{2D_x} w_{i'} (g(\chi_{i'}) \! - \! \mu_t)(g(\chi_{i'}) \! - \! \mu_t)^{\intercal}  +  R \nonumber \\
        &- \Big(\sum_{i=0}^{2D_x} w_i (\chi_i \! - \! \bar{m}_t)(g(\chi_i) \! - \! \mu_t)^{\intercal} \Big)^\intercal \bar{S}_t^{-1} \Big( \sum_{j=0}^{2D_x} w_j (\chi_j \! - \! \bar{m}_t)(g(\chi_j) \! - \! \mu_t)^{\intercal} \Big) \Big| \, .
    \end{align}
\end{proof}
The second term can be re-arranged to:
\begin{align} 
    \Big(\sum_{i=0}^{2D_x} w_i (\chi_i \! - \! \bar{m}_t)(g(\chi_i) \! - \! \mu_t)^{\intercal} \Big)^\intercal \bar{S}_t^{-1} \Big( \sum_{j=0}^{2D_x} w_j (\chi_j \! - \! \bar{m}_t)(g(\chi_j) \! - \! \mu_t)^{\intercal} \Big) \nonumber \\
    = \sum_{i=0}^{2D_x} \sum_{j=0}^{2D_x} w_i (g(\chi_i) \! - \! \mu_t)(\chi_i \! - \! \bar{m}_t)^{\intercal} \bar{S}_t^{-1} w_j (\chi_j \! - \! \bar{m}_t)(g(\chi_j) \! - \! \mu_t)^{\intercal} \, . \label{eq:proof_ut02}
\end{align}
%
%
Note that for $j=0$, $(\chi_j \! - \! \bar{m}_t) \! = \! (\bar{m}_t \! - \! \bar{m}_t) \! = \! 0$. Let $D_\lambda = D_x + \lambda$. For $j\geq1$:
\begin{align} \label{eq:proof_ut03}
    &(\chi_i \! - \! \bar{m}_t)^{\intercal} \bar{S}_t^{-1} w_j (\chi_j \! - \! \bar{m}_t) \\
    &\ = (\bar{m}_t \! + \! (-1)^{i} \sqrt{D_\lambda} \big[ \sqrt{\bar{S}_t} \big]_i \! - \! \bar{m}_t)^{\intercal} \bar{S}_t^{-1} \frac{1}{D_\lambda} (\bar{m}_t \! + \! (-1)^{j} \sqrt{D_\lambda} \big[ \sqrt{\bar{S}_t} \big]_j \! - \! \bar{m}_t) \nonumber \\
    &\ = \frac{1}{D_\lambda} (\sqrt{D_\lambda})^{2} (-1)^{i+j} \big[ \sqrt{\bar{S}_t} \big]_i^{\intercal} \bar{S}_t^{-1} \big[ \sqrt{\bar{S}_t} \big]_j  \\
    &\ = (-1)^{i+j} \big[ \sqrt{\bar{S}_t} \big]_i^{\intercal} \bar{S}_t^{-1} \big[ \sqrt{\bar{S}_t} \big]_j \, . \label{eq:proof_ut04}
\end{align}
%
Furthermore, note that column selection $\big[\cdot\big]_i$ is equivalent to right-multiplication with a canonical basis vector $e_i$;
\begin{align} \label{eq:proof_ut05}
    \big[ \sqrt{\bar{S}_t} \big]_i^{\intercal} \bar{S}_t^{-1} \big[ \sqrt{\bar{S}_t} \big]_j = \big(\sqrt{\bar{S}_t} e_i \big)^{\intercal} \bar{S}_t^{-1} \big( \sqrt{\bar{S}_t} e_j \big) =e_i^{\intercal} \sqrt{\bar{S}_t}^{\intercal} \bar{S}_t^{-1}  \sqrt{\bar{S}_t} e_j \, .
\end{align}
Since $\bar{S}_t$ is a normal matrix, the eigendecomposition $\bar{S}_t = V \Omega V^{-1}$ generates an orthonormal eigenvector matrix $V$, implying $V^{-1} = V^{\top}$, and a diagonal matrix of eigenvalues $\Omega$. This means that $\sqrt{\bar{S}_t} = V \Omega^{1/2} V^{-1}$, and that:
\begin{align}
    \sqrt{\bar{S}_t}^{\intercal} \bar{S}_t^{-1} \sqrt{\bar{S}_t} &= \big( V \Omega^{1/2} V^{-1} \big)^{\intercal} V \Omega^{-1} V^{-1} \big( V \Omega^{1/2} V^{-1} \big) \\
    &= V \Omega^{1/2} V^{-1} V \Omega^{-1} V^{-1} V \Omega^{1/2} V^{-1} \\
    &= V V^{-1} = I.
\end{align}
Therefore, $e_i^{\intercal} I e_j$ will be $1$ for all $i = j$ and $0$ for $i \neq j$. We can thus identify two cases in the double sum in \eqref{eq:proof_ut02}, one of which is always $0$:
\begin{align}
     \sum_{i=0}^{2D_x} \sum_{j=0}^{2D_x} &w_i (g(\chi_i) \! - \! \mu_t)(\chi_i \! - \! \bar{m}_t)^{\intercal} \bar{S}_t^{-1} w_j (\chi_j \! - \! \bar{m}_t)(g(\chi_j) \! - \! \mu_t)^{\intercal} \\
     &=  \sum_{i=0}^{2D_x} \sum_{j=i} w_i (g(\chi_i) \! - \! \mu_t)(-1)^{(i+j)} \, 1 \, (g(\chi_j) \! - \! \mu_t)^{\intercal} \nonumber \\
     &\qquad + \sum_{i=0}^{2D_x} \sum_{j \neq i} w_i (g(\chi_i) \! - \! \mu_t)(-1)^{(i+j)} \, 0 \, (g(\chi_j) \! - \! \mu_t)^{\intercal} \\
     &= \sum_{i=0}^{2D_x} w_i (g(\chi_i) \! - \! \mu_t) (g(\chi_i) \! - \! \mu_t)^{\intercal} \, ,
\end{align}
where the $(-1)^{(i+j)}$ drops out because for $i=j$, $i+j$ will always be even.
%
%
One may now recognize that \eqref{eq:proof_ut01} has two terms that cancel each other:
\begin{align}
    \frac{1}{2}\ln | \sum_{i'=0}^{2D_x} w_{i'} (g(\chi_{i'}) \! - \!  \mu_t)(g(\chi_{i'}) \!  - \!  \mu_t)^{\intercal} \! + \! R \! - \! \sum_{i=0}^{2D_x} w_i & (g(\chi_i) \! - \! \mu_t)(g(\chi_i) \! - \! \mu_t)^{\intercal} | \nonumber \\ 
    &=  \frac{1}{2} \ln |R| \, .
\end{align}
Using this result and Lemma 1, we have proven Theorem 3.

%
\bibliographystyle{splncs04}
\bibliography{references}

\end{document}